\documentclass[11pt,twoside]{article}

\usepackage{asp2006}
\usepackage{epsf}
\usepackage{psfig}
\usepackage{lscape}
\usepackage{graphicx}

\begin{document}
\title{Modelling the flaring emission at the Galactic Centre}   
\author{E. M. Howard}   

\begin{abstract} 
The massive black hole at the Galactic Centre is known to be variable
in radio, millimeter, near-IR and X-rays. 
We investigate the physical processes responsible for the variable observed emissions from the compact radio source Sgr A*. We study the evolution of the variable emission region and present light curves and time-resolved spectra of emissions from the accretion disk, close to the event horizon, near the marginally stable orbit of a Kerr black hole. 
\end{abstract}


\section{Outline}

We study a complete system composed of a rotating black hole, a thin accretion disc and a co-rotating spot within the disk. The co-rotating accretion disc is geometrically thin and optically thick, therefore we take into account only photons coming from the equatorial plane to the observer. (See \citet{fabian00}) We use a ray tracing code to determine how the relativistic effects such as beaming, Doppler shift and gravitational lensing influence the Sgr A* emission as it propagates from a co-moving frame, to a distant observer, conceived as an ideal array of pixels. (See \citet{kato98}).

\section{Model used}

The model  allows us to:

(i) study the emission from the plunge region, and

(ii) specify a time dependent form of intrinsic emissivity as a function of polar coordinates in the disk plane.

We study the plunge region of an in-falling spot, located between the event horizon and the last stable orbit, and   look for any significant contribution to the observed emission originating from within the localized region. (See \citet{dovciak04a})
The spot is orbiting within the the equatorial plane of a Kerr black hole. The radius of the spot orbit is spin dependent. The transfer of photons is computed by integration of the geodesic equation. 
The model used calculates the local disc emission in polar coordinates, enabling us to choose a non-axisymmetric area of integration and to handle emission from a localized spot within the disk. (See \citet{dovciak04}) The spot is a single region of isotropic, monochromatic emission following null geodesic trajectories. Intrinsic emissivity fades gradually with the distance from the spot centre 
and it is not constant in time. Intrinsic emissivity varies according to a normal distribution
and is a function of the axial position of the spot within the disk and indirectly, of time. 

\begin{figure}
\begin{center}
\includegraphics[width=10cm]{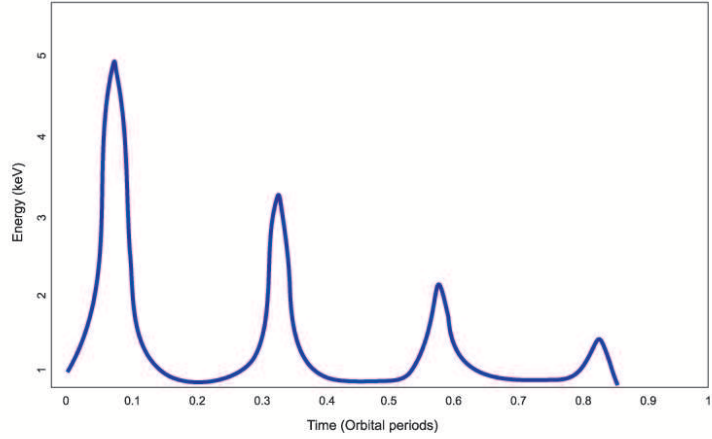}
\end{center}
\caption{Centroid energy as function of the orbital phase of an in-falling spot near the horizon of an extreme Kerr black hole. Time is expressed in orbital periods.}\label{fig1}
\end{figure}

\begin{figure}
\begin{center}
\includegraphics[width=10cm]{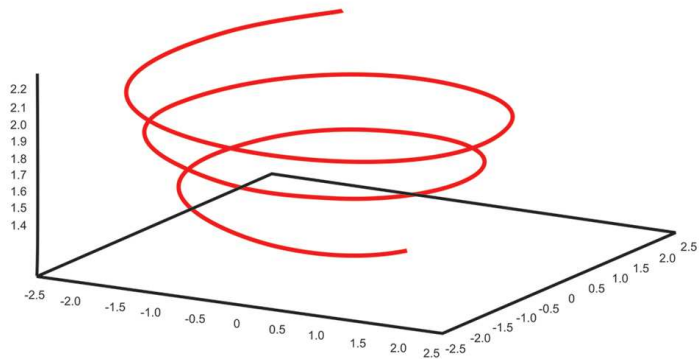}
\end{center}
\caption{Plunging trajectory of a photon into the event horizon of an extreme Kerr black hole (in Schwarzschild radius units)}\label{fig2}
\end{figure}

\section{Results}
By analyzing dynamical spectra, we address the question of whether existing data is consistent with a simple orbiting or in-falling spot model (See Fig. 1), in which the presence of relativistic effects play a significant role.
We fit various parameters such as black-hole angular momentum, observer inclination angle relative to the disk axis and the size and shape of the emission area. We analyze light curves (See Fig. 2) for different inclination angles, spin values, brightness  of the spot, initial phases of the spot on the orbit and orientation angles of the equatorial plane relative to the observer's sky plane. Our results show a dependence between the intrinsic emissivity model and the observed flux, suggesting a correlation between the energy spectral index and the characteristic profile of the observed light curve. 
The results show also that most rapid variability occurs at the marginally stable orbit. The variation of most
of the observed spectral features is higher for orbits close to the event horizon and it is also direct proportional with the inclination angle.




\end{document}